\documentclass[12pt,preprint]{aastex}

\def\ll{~$\lambda\lambda$}

\def\kms{km\,s$^{-1}$}

\def\msun{M$_{\odot}$}

\def\na{{\it n/a}}
\def\gtrsim{\mathrel{\hbox{\rlap{\hbox{\lower3pt\hbox{$\sim$}}}\hbox{\raise2pt\hbox{$>$}}}}}
\def\lesssim{\mathrel{\hbox{\rlap{\hbox{\lower3pt\hbox{$\sim$}}}\hbox{\raise2pt\hbox{$<$}}}}}

\shorttitle{The massive overcontact binary VFTS\,352}
\shortauthors{Almeida et al.}

\begin{document}


\title{
Discovery of the massive overcontact binary VFTS\,352: Evidence for enhanced internal mixing
}


\author{L.A. Almeida\altaffilmark{1,2,3},
H. Sana\altaffilmark{4,5},
S.E. de Mink\altaffilmark{6},
F. Tramper\altaffilmark{6},
I. Soszy{\'n}ski\altaffilmark{7},
N. Langer\altaffilmark{8},
R.H. Barb\'a\altaffilmark{9},
M. Cantiello\altaffilmark{10}
A. Damineli\altaffilmark{2},
A. de Koter\altaffilmark{6,11},
M. Garcia\altaffilmark{12},
G. Gr\"afener\altaffilmark{16},
A. Herrero\altaffilmark{13,14},
I. Howarth\altaffilmark{15},
J. Ma\'{i}z Apell\'aniz\altaffilmark{12},
C. Norman\altaffilmark{1},
O.H. Ram\'{i}rez-Agudelo\altaffilmark{6},
J.S. Vink\altaffilmark{16}
}
\altaffiltext{1}{Department of Physics \& Astronomy, Johns Hopkins University, Bloomberg Center for Physics and Astronomy, Room 520, 3400 N Charles St}
\altaffiltext{2}{Instituto de Astronomia, Geof\'isica e Ci\^encias, Rua do Mat\~ao 1226, Cidade Universit\'aria S\~ao Paulo, SP, Brasil, 05508-090}
\altaffiltext{3}{Email: {\tt leonadodealmeida.andrade@gmail.com}}
\altaffiltext{4}{European Space Agency/Space Telescope Science Institute, 3700 San Martin Drive, Baltimore, MD 21218, United States}
\altaffiltext{5}{Institute of Astronomy, KU Leuven, Celestijnlaan 200D, 3001 Leuven, Belgium}
\altaffiltext{6}{Anton Pannenkoek Astronomical Institute, University of Amsterdam, 1090 GE Amsterdam, The Netherlands}
\altaffiltext{7}{Warsaw University Observatory, Al. Ujazdowskie 4, 00-478 Warszawa, Poland}
\altaffiltext{8}{Argelander-Institut für Astronomie, 	
	der Universit{\"a}t Bonn, 	
	Auf dem Hügel 71, 	
	53121 Bonn, 	
	Germany}
\altaffiltext{9}{Departamento de F\'{\i}sica y Astronom\'{\i}a, Universidad de La Serena, Av. Cisternas 1200 Norte, La Serena, Chile}
\altaffiltext{10}{Kavli Institute for Theoretical Physics, University of California, Santa Barbara, CA 93106, USA}
\altaffiltext{11}{Instituut voor Sterrenkunde, 
           Universiteit Leuven, 
           Celestijnenlaan 200 D, 
           3001, Leuven, Belgium}
\altaffiltext{12}{Departamento de Astrof\'isica, Centro de Astrobiolog\'ia (INTA-CSIC), campus ESA, apartado postal 78, 28 691 Villanueva de la Ca\~nada, Madrid, Spain}
\altaffiltext{13}{Instituto de Astrof\'{i}sica de Canarias, 
           C/ V\'{i}a L\'{a}ctea s/n, E-38200 La Laguna, Tenerife,
           Spain}
\altaffiltext{14}{Departamento de Astrof\'{i}sica, 
           Universidad de La Laguna, 
           Avda. Astrof\'{i}sico Francisco S\'{a}nchez s/n, 
           E-38071 La Laguna, Tenerife, Spain}
\altaffiltext{15}{Department Physics \& Astronomy, University College London, Gower Street, London WC1E 6BT, UK}
\altaffiltext{16}{Armagh Observatory, College Hill, Armagh, BT61 9DG, Northern Ireland, UK}



\begin{abstract}
The contact phase expected to precede the coalescence of two massive stars is poorly 
characterized due to the paucity of observational constraints. Here we report on the 
discovery of VFTS\,352, an O-type binary in the 30 Doradus region, as the most massive 
and earliest spectral type overcontact system known to date. We derived the 3D geometry 
of the system, its orbital period $P_{\rm orb}=1.1241452(4)$~d, components' effective 
temperatures -- $T_1=42\,540\pm280$~K and $T_2=41\,120\pm290$~K -- and dynamical 
masses -- $M_1=28.63\pm0.30~M_{\odot}$ and $M_2=28.85\pm0.30~M_{\odot}$.
Compared to single-star evolutionary models, the VFTS\,352 components are too hot 
for their dynamical masses by about 2700 and 1100~K, respectively. These results can be 
explained naturally as a result of enhanced mixing, theoretically predicted to occur 
in very short-period tidally-locked systems. The VFTS\,352 components are two of the 
best candidates identified so far to undergo this so-called chemically homogeneous 
evolution. The future of VFTS\,352 is uncertain. If the two stars merge, a very rapidly 
rotating star will be produced. Instead, if the stars continue to evolve homogeneously 
and keep shrinking within their Roche Lobes, coalescence can be avoided. In this case, 
tides may counteract the spin down by winds such that the VFTS\,352 components may, at the 
end of their life, fulfill the requirements for long gamma-ray burst (GRB) progenitors in 
the collapsar scenario. Independently of whether the VFTS\,352 components become GRB 
progenitors, this scenario makes VFTS\,352 interesting as a progenitor of a black 
hole binary, hence as a potential gravitational wave source through black hole-black hole 
merger.

\end{abstract}


\keywords{stars: early-type --- stars: massive --- binaries: spectroscopic --- 
binaries: eclipsing --- binaries: close ---  stars: individual: VFTS\,352}

\section{INTRODUCTION}\label{sect:intro}
Massive stars are one of the most important cosmic engines driving the evolution 
of galaxies throughout the history of the universe. In the Milky Way, most massive 
stars are born as part of a binary system with an orbital period $P_{\rm orb} < 4$~yr 
\citep{MHG09, SdMdK12}, i.e.\ close enough to interact during their lifetime through 
mass-exchange or coalescence \citep{PJH92}. Such binary interactions have important 
consequences for the subsequent evolution of the components of the system and lead to 
key astrophysical phenomena such as double compact binaries, hydrogen-deficient core-collapse 
supernovae and gamma-ray bursts (GRBs). The high frequency of close binaries further challenges 
the accepted predominance of the single-star evolutionary channel. 

Binary interaction may happen early in the evolution of the stars. According to the massive 
star multiplicity properties derived from galactic young open clusters \citep{SdMdK12}, 
40\% of all O-type stars may interact with a nearby companion before leaving the main 
sequence and over half these main sequence interactions (about a quarter of all stars born 
as O type) will lead to a deep contact phase of which the most likely outcome will be 
coalescence. Such a process modifies the mass function of massive star clusters, and may 
help to create some of the most massive stars known \citep{CSH10, SIdM14}.

Despite possibly affecting almost one quarter of all O-type stars, the deep contact phase 
is one of the least understood phases of massive binary evolution. This results from the 
complex interplay between a variety of physical processes, including mass exchange, tidal 
locking, intense mutual illumination, internal mixing, angular momentum loss, as well as 
possible adjustments to the stars internal structure \citep{Pol94, WLB01, deMink+2007}.

Our limited understanding also stems from the almost complete absence of observational 
constraints. Observationally, these interacting binaries are expected to reveal themselves 
as {\it overcontact} binaries, i.e.\ an orbital configuration in which both components are 
overfilling their Roche lobes \citep{Wilson2001}. However, only three O-type overcontact 
binaries are known to date (see Sect.~\ref{sec:overfilling}).

\begin{figure}
 \resizebox{\hsize}{!}{\includegraphics[angle=0]{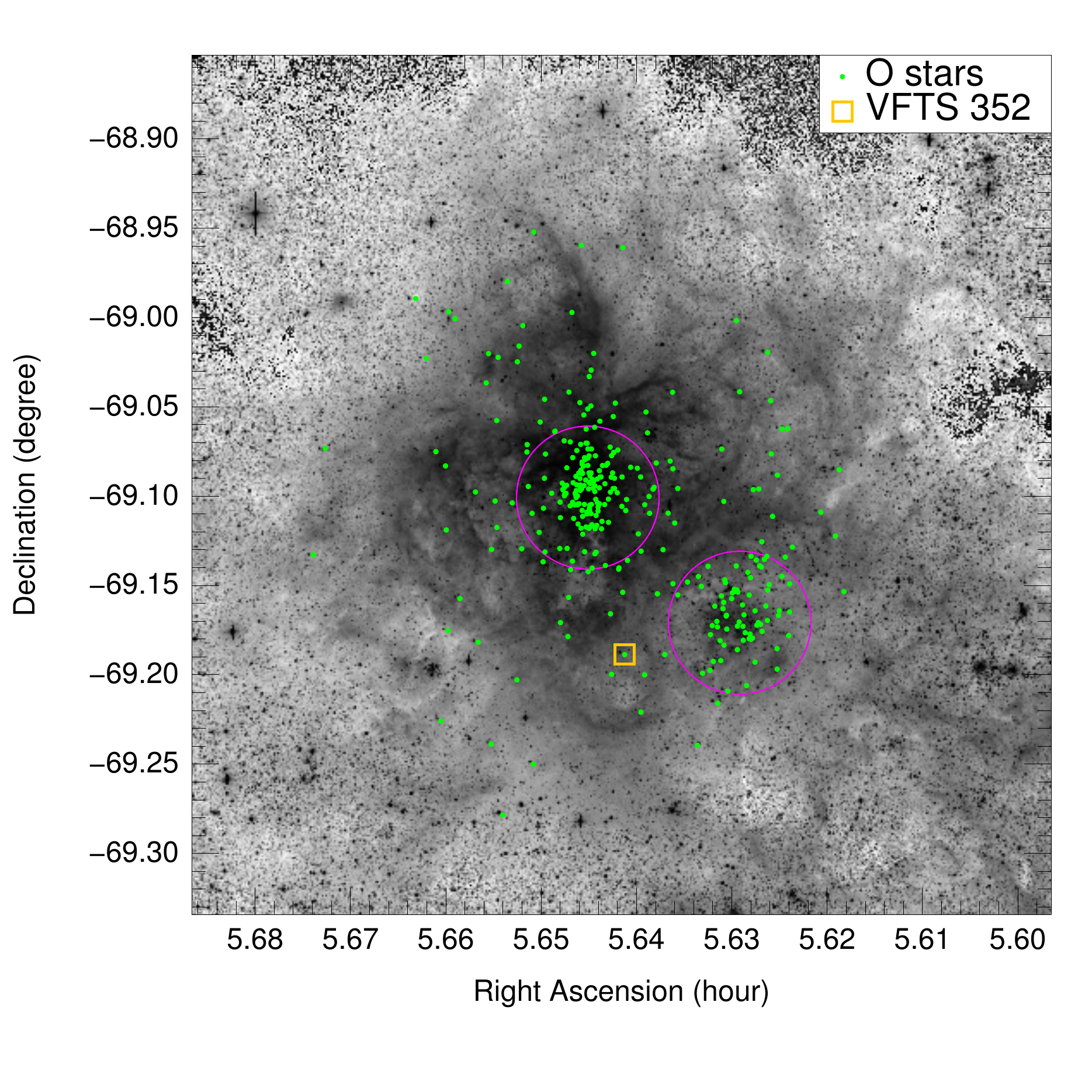}}
 \caption{Location of VFTS\,352 in the 30 Dor field-of-view. All O-type stars observed by the 
 VFTS are also marked. The two circles indicate the locations of NGC2070 (at the center) and 
 NGC 2060 (to the South-West). 
}
 \label{fig:field}
\end{figure}

Located in the Tarantula Nebula (30 Dor, Fig.~\ref{fig:field}) in the Large Magellanic 
Cloud (LMC), VFTS\,352 is a double-lined spectroscopic binary identified by the VLT-FLAMES 
Tarantula Survey \citep[VFTS,][]{ETHB11,SdKdM13}. Classified as 
O4.5 V(n)((fc))z + O5.5 V(n)((fc))z \citep{WSSD14}, both companions present rotationally 
broadened \ion{H}{1}, \ion{He}{1} and \ion{He}{2} absorption lines. In this study we present 
the results of the analysis of an 18-month spectroscopic monitoring of VFTS\,352 combined 
with $\sim$12 years of OGLE-III and IV photometry \citep[]{Udalski2008Aca}.

\section{OBSERVATIONS AND DATA REDUCTION}\label{observation}

\subsection{Spectroscopy}\label{sec:spectroscopy}
New spectroscopic data of VFTS\,352 were collected as part of the Tarantula Massive Binary 
Monitoring program (PI: Sana) at the ESO Very Large Telescope (VLT). 32 spectroscopic epochs 
were obtained from October 2012 to March 2014 using the FLAMES-GIRAFFE multi-object 
spectrograph and the LR02 grism, providing continuous coverage of the 3950--4550~\AA\ region 
with a spectral resolving power $\lambda / \Delta \lambda$ of 6,400. Three sets of back-to-back 
exposures were taken to ensure optimal cosmic-ray removal. The six LR02 spectra obtained as 
part of the VFTS were further included in our analysis.

\begin{figure*}
 \includegraphics[width=\textwidth,height=6cm,angle=0]{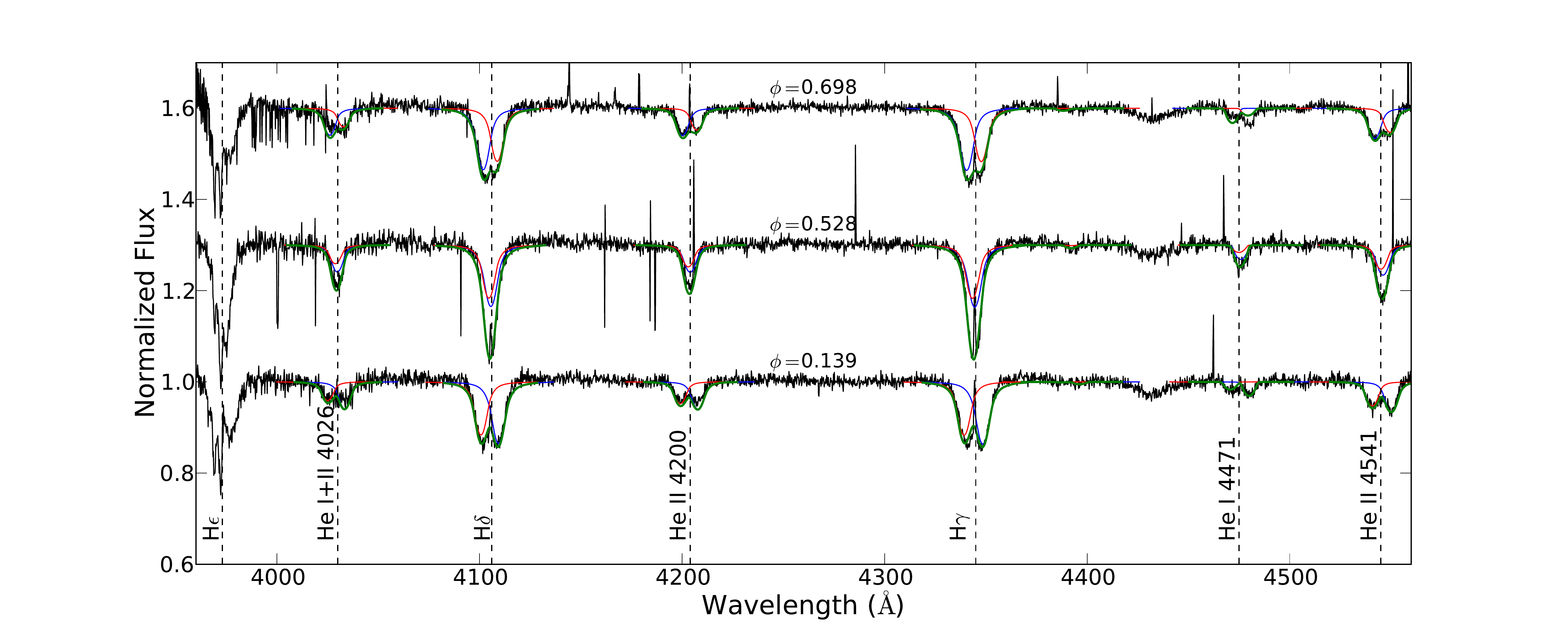}
 \caption{Examples of VFTS\,352 spectra at three different epochs illustrating the Doppler shift of the \ion{H}{1}, \ion{He}{1}, and \ion{He}{2} lines. Overplotted are the {\sc fastwind} models described in Section~\ref{sec:atm_models} for primary (blue lines), secondary (red lines), and the combination of both components (green lines). The combined line profiles reproduce the observed binary spectrum very well.}
\label{fig:spectrum}
\end{figure*}

The data reduction was performed using the ESO CPL GIRAFFE pipeline v.2.12.1. It involved the 
standard steps of bias subtraction, flat-field structure removal, extraction, and wavelength 
calibration. Additional steps of sky correction and normalization were applied following the 
procedures described in \citet{ETHB11} and \citet{SdKdM13}, respectively. 
Figure~\ref{fig:spectrum} shows examples of normalized individual spectra.

\subsection{Photometry}

$V-$ and $I-$band time-resolved photometry were obtained by the OGLE-III and IV projects 
\citep{Udalski2008Aca}. The temporal coverage of the OGLE-III and IV data is of $\sim$7.5 
and 5.2 years, respectively, with an accuracy of a few mmag. There is a small shift 
($\sim$0.01 mag) in brightness between OGLE-III and OGLE-IV light curves due to independent 
calibration which has an accurancy of about 0.02 mag \citep{Udalski2008Aca}. The total number 
of VFTS\,352 measurements available in the $V-$ and $I-$bands  is 90 and 760. Both light 
curves  show ellipsoidal variations with semi-amplitude of $\sim 0.1$ mag indicating that at 
least one component of the system is filling its Roche lobe or is close to doing so.

\section{ANALYSIS AND RESULTS}\label{analysis}

\subsection{Radial velocity measurements}
\label{sec:rvsolution}

The radial velocities  of the VFTS\,352 components were obtained  following the procedure 
described in \citet{SdKdM13}. In brief, we simultaneously adjusted the \ion{He}{1}\ll{4388}, 
4471 and \ion{He}{2}\ll{4200}, 4541 lines at all epochs using Gaussian profiles. The shape of 
the Gaussian profile for each spectral line and each component is kept constant at all epochs, 
allowing for enhanced robustness of the fitting at phases where the lines are heavily blended 
or where the $S/N$ is poorer. The obtained RVs and 1$\sigma$ error bars are available at the 
Centre de Donn\'ees astronomiques de Strasbourg\footnote{{\tt http://vizier.u-strasbg.fr/}} 
(CDS) together with the journal of the observations. The best-fit RV-only orbital solution 
yields semi-amplitudes of $K_1=324.9\pm5.8~\rm km\,s^{-1}$ and $K_2 =315.6\pm6.2~\rm km\,s^{-1}$ 
for the primary and secondary components. In the next section, the RV-only solution that we 
obtain is used as a starting point for the combined photometric and spectroscopic orbital 
solution.

\subsection{Simultaneous modeling of light curves and radial velocities}\label{sec:wdc}

To obtain the geometrical and physical parameters of VFTS\,352, we simultaneously adjusted 
synthetic light and radial velocity curves generated by the Wilson Devinney 
code \citep[WDC,][]{WiD71} to the $V-$ and $I-$band OGLE-IV light curves and the VLT/FLAMES 
radial velocity curves. We follow the procedure presented in \citet{2012MNRAS.423.478A} to 
search for the global solution. In essence, the WDC is used as a \emph{function} to be 
optimized by the genetic algorithm {\tt PIKAIA} \citep{1995ApJS.101.309C} followed by a 
Markov chain Monte Carlo (MCMC) procedure \citep{Gilks1996} to sample the parameters and 
obtain their uncertainties.

\begin{figure}
 \center
 \includegraphics[bb=190 10 590 590]{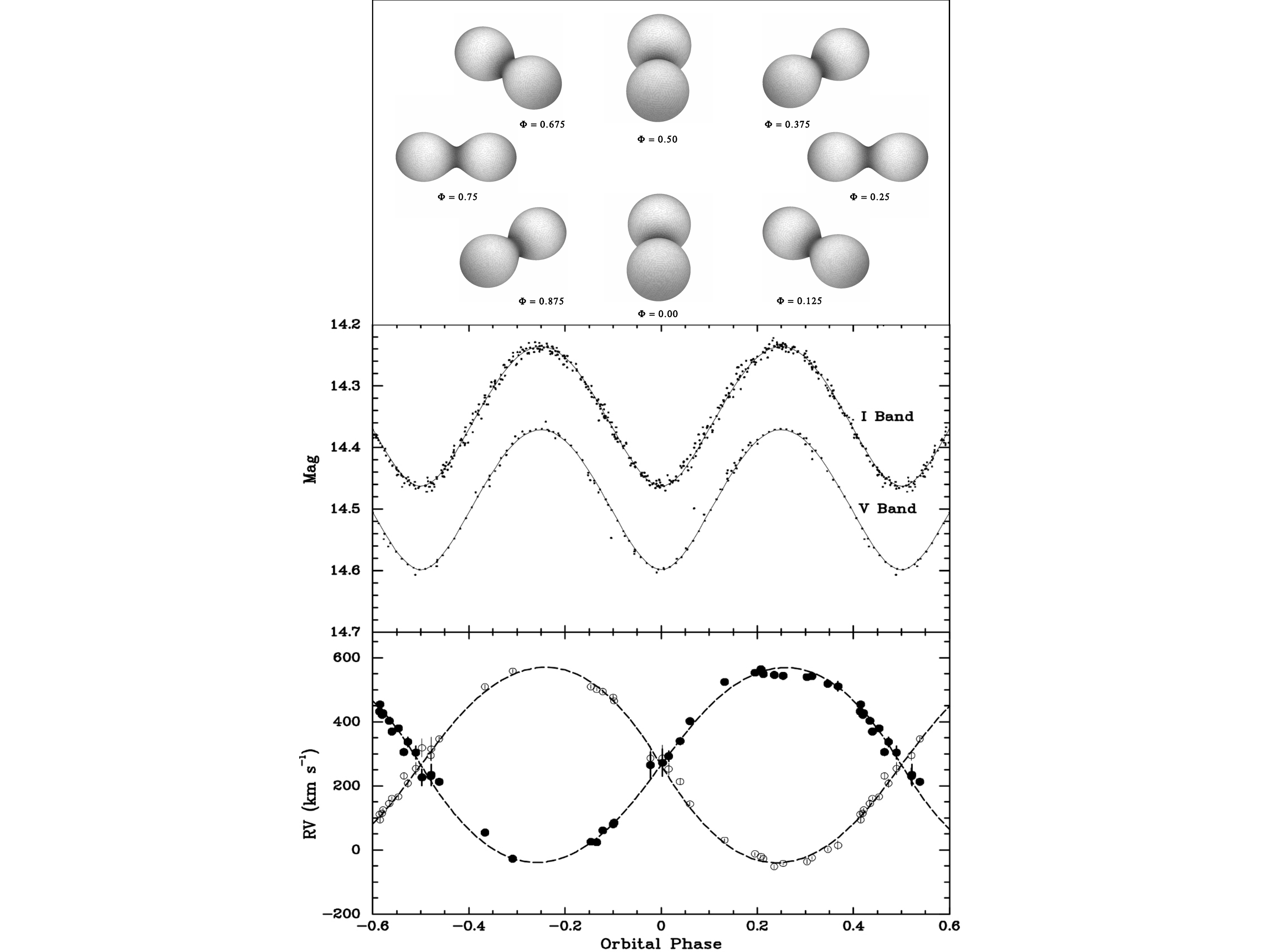}
 \caption{From the top to the bottom panels are: snapshots of VFTS\,352 at few orbital 
 phases, as viewed from an orbital inclination angle $i$ of 55.6 deg; and best WDC model 
 simultaneously fitted to the $V-$ and $I-$ band light curves and radial velocity curves, 
 respectively. 
 }
 \label{lc_rv}
\end{figure}

To restrict the size of the explored parameter space, we use priors obtained from our 
initial radial velocity solution (Sect.~\ref{sec:rvsolution}) and from the known spectral 
type of the components \citep{WSSD14}. {\rm We constrain the mass ratio ($q=M_2 / M_1$) 
between $0.95 < q < 1.05$, and effective temperatures, $40900<T_{\rm eff,1}<44900$ and 
$38900~K<T_{\rm eff,2}<42900~K$},  with 42900 and 40900 K \citep{WSSD14} as a starting point 
for the primary and secondary respectively. We also keep fixed the distance $d = 50$~kpc and 
the $V-$band extinction $A_v = 1.109\pm0046$ \citep{Maiz+2014}. As the light curves show 
ellipsoidal variation and $q \sim 1$, we try to model the data with 
overcontact, contact, and semi-detached configurations, but only the overcontact model 
provides a plausible solution (see the top panel of Figure~\ref{lc_rv}). In all these models, 
non-Keplerian effects (tidal distortion of the stars) are taken into account.

To  account for the temperature difference between the components 
(see Section~\ref{sec:atm_models}), we use the mode 3 of the WDC which has no
constrain in the temperature, bolometric albedo, gravity brightening, and limb 
darkening coefficients of the secondary star. WDC converts the mean surface effective 
temperatures, $T_1$ and $T_2$, in polar temperature using Eq. 8 of \citet{Wilson1979} and 
then computes the local temperature using gravity brightening law and polar temperature. 
The linear limb darkening coefficients were computed for both stars as a function of 
$T_{\rm eff}$ and $\log g$. In this purpose, the coefficients calculated by \citet{Diaz1995} 
and \citet{Claret1995} are automatically updated while the program is running. The best 
solution for VFTS\,352 is shown in Figure~\ref{lc_rv} and the adjusted and derived parameters 
are listed in Table~\ref{system:parameters}.

\begin{table}
\begin{small}
\begin{center}
\caption{Best fit parameters resulting from the simultaneous adjustment in the $I-$ and $V-$ band light curves and radial velocity curves of the components.}         
\label{system:parameters}    
\begin{tabular}{l r}        
\hline
\hline                 
 Parameter & Value \\     
\hline
\multicolumn{2}{c}{Adjusted Parameters}   \\
\hline
$P_{\rm orb}$ (d)   &     $1.1241452 \pm 0.0000004$ \\
$T_0$ (HJD)         &     $2455261.119 \pm 0.003$   \\
$q=M_2/M_1$         &      $1.010 \pm 0.010$        \\
$\Omega^a_1$        &      $3.584 \pm 0.014$          \\
$\Omega^a_2$        &      $3.584 \pm 0.014$          \\
$T_1$ (K)           &      $42540 \pm 280$         \\
$T_2$ (K)           &      $41120 \pm 290$         \\
$i$ ($^{\circ}$)    &      $55.60 \pm 0.20$          \\
$a^b$ (R$_{\odot}$) &      $17.55 \pm 0.06$          \\
$\gamma$ (\kms)     &      $262.8 \pm 1.2$          \\
\hline
\multicolumn{2}{c}{Roche radii} \\
\hline
$R_1^{\rm pole}/R_\mathrm{L}$   &      $1.005 \pm 0.001$      \\
$R_1^{\rm side}/R_\mathrm{L}$   &      $1.070 \pm 0.001$      \\
$R_1^{\rm back}/R_\mathrm{L}$   &      $1.192 \pm 0.001$      \\
$R_1^{\rm mean}/R_\mathrm{L}$   &      $1.089 \pm 0.001$      \\
$R_2^{\rm pole}/R_\mathrm{L}$   &      $1.004 \pm 0.001$      \\
$R_2^{\rm side}/R_\mathrm{L}$   &      $1.067 \pm 0.001$      \\ 
$R_2^{\rm back}/R_\mathrm{L}$   &      $1.189 \pm 0.001$      \\
$R_2^{\rm mean}/R_\mathrm{L}$   &      $1.087 \pm 0.001$      \\
\hline
\multicolumn{2}{c}{Derived parameters} \\
\hline
$M_1$ (M$_{\odot}$)     &  $28.63 \pm 0.30$    \\
$M_2$ (M$_{\odot}$)     &  $28.85 \pm 0.30$    \\
$R^\mathrm{mean}_1$ (R$_{\odot}$)   &  $7.22 \pm 0.02$    \\
$R^\mathrm{mean}_2$ (R$_{\odot}$)   &  $7.25 \pm 0.02$    \\
$\log g_1$ [cm s$^{-2}$]   &  $4.18 \pm 0.01$    \\
$\log g_2$ [cm s$^{-2}$]   &  $4.18 \pm 0.01$   \\
$M_1^{\rm bol}$         &  $-8.26 \pm 0.01$   \\
$M_2^{\rm bol}$         &  $-8.08 \pm 0.01$   \\
\hline                   
\end{tabular} \\
\end{center}
$^a$ Roche surface potential - $\Omega_2$ = $\Omega_1$ in mode 3 of the WDC; \\
$^b$ Binary separation.
\end{small}
\end{table}

\subsection{Atmosphere analysis}\label{sec:atm_models}

We employ the line-blanketed non-local thermodynamic equilibrium atmosphere code {\sc fastwind} 
\citep{Puls2005} to model both components of VFTS\,352. For our initial models we use the 
values for the radius, gravity and temperatures listed in Table~\ref{system:parameters} as 
input. We adopt the \citet{vink2001} mass-loss rates based on the stellar parameters and an 
LMC metallicity and we use the scaling with the escape velocity $(v_{\mathrm{esc}})$ to 
estimate the wind terminal velocity 
\citep[$v_{\infty} = 2.65 \times v_{\mathrm{esc}}$;][]{Kudritzki2000}. Finally, we adopt 
rotation rates corresponding to a fully synchronized system, i.e.\ $v_\mathrm{rot} = 325$~\kms, 
and we use the orbital inclination value ($i$) from Table~\ref{system:parameters} to compute 
the projected rotation velocity at the equator.

The model that we obtained (Figure~\ref{fig:spectrum}) further takes into account the phase 
dependance of the brightness ratio of the two components and provides a satisfactory 
representation of the  data. In addition to the physical parameters, we also investigate the 
helium surface abundance in the VFTS\,352 spectra. While we could not find any clear signature 
of enrichment, we estimate that a surface enrichment corresponding to a mass fraction 
$X_\mathrm{He}$ less than 40\%\ would be hard to detect with our data. 

Figure~\ref{fig:spectrum} also reveals small variations in the intensity of the \ion{He}{1} 
lines. Such a behaviour is reminiscent of the Struve-Shade effect 
\citep{Bagnuolo1999, Linder2007} and possibly results from  temperature structure on the 
(deformed) stellar surface \citep{PalateRauw2012}. Future spectral disentangling, taking 
into account the phase dependance flux-ratio of the object -- may help to lower the present 
upper limit on the He surface abundance. Combined with   phase-resolved spectral modelling, 
it would further help to elucidate the nature of the \ion{He}{1} lines line profile 
variability. These two aspects however lay beyond the scope of the present work.

\subsection{Orbital period variation}\label{sec:per}

To search for indication of variations in the orbital configuration of VFTS\,352, we 
scrutinize its orbital period. We split the OGLE-III and IV $I-$band photometric data -- 
which span 12.5 years -- into 15 parts that we analyze separately using the WDC. As all the 
VFTS\,352 parameters are well determined (see Table~\ref{system:parameters}), we allow them 
to vary only within $\pm 1~\sigma$ from their best-fit value and we leave the orbital period 
as the only entirely free parameter. The MCMC procedure was used to sample the posterior 
distribution of the orbital period and obtain the best value as well as its uncertainty.

\begin{figure}
 \center
 \includegraphics[width=10cm,angle=-90]{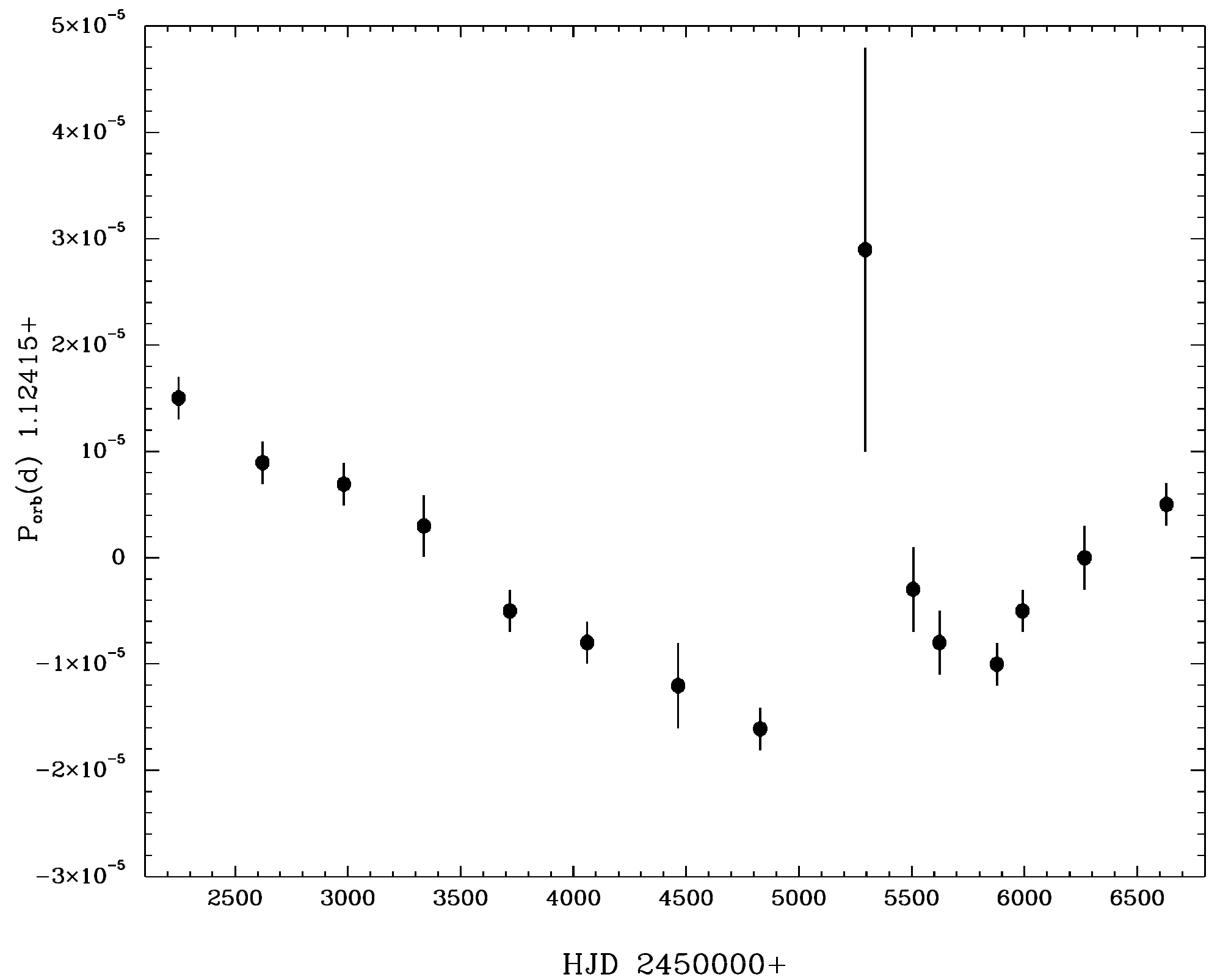}
 \caption{Orbital period variation of VFTS\,352. The orbital period measurements obtained 
 from OGLE-III and IV $I-$band photometric data are shown as filled and open circles 
 respectively.}
 \label{fig:oc}
\end{figure}

With a peak-to-peak variation of $\sim$2~s over a 12.5~yr time span, the variation time 
scale of the VFTS\,352 orbital period is approximately $P/\dot{P} \approx 0.6$ Myr, 
see Fig.~\ref{fig:oc}.  Such a relatively long time scale suggests that the system may 
currenty be stable, thus the overcontact configuration relatively long lived. In that case, 
the relevant evolutionary time scale for the system is the nuclear time scale. The apparent 
stability of the VFTS\,352 configuration would then favour an evolutionary stage that 
corresponds to the rather long-lasting overcontact phase predicted by Case A mass transfer 
scenarios \citep{Pol94, WLB01}.

\section{DISCUSSION AND CONCLUSION}
\label{discussion}

\begin{figure*}
 \includegraphics[width=8cm]{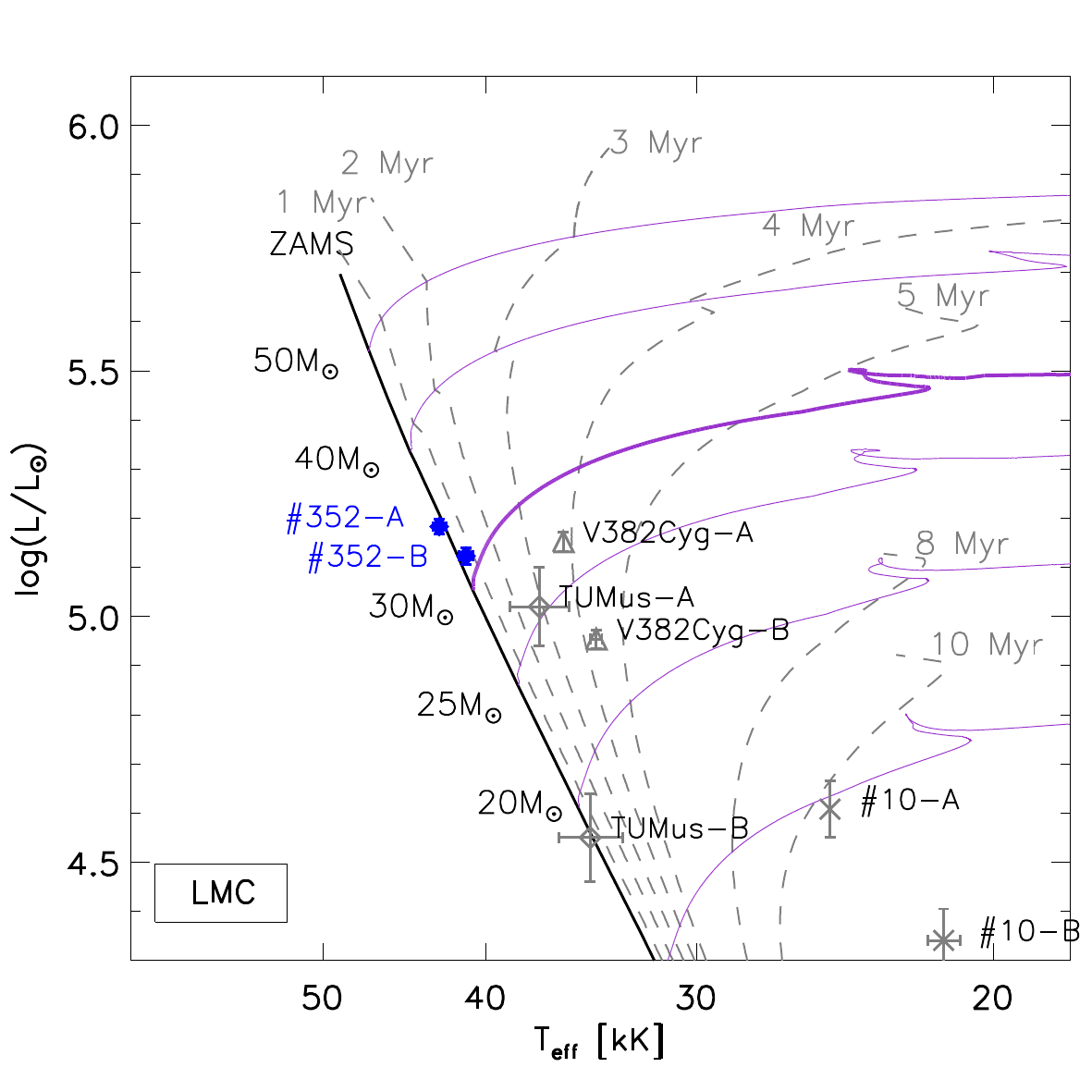}
 \includegraphics[width=8cm]{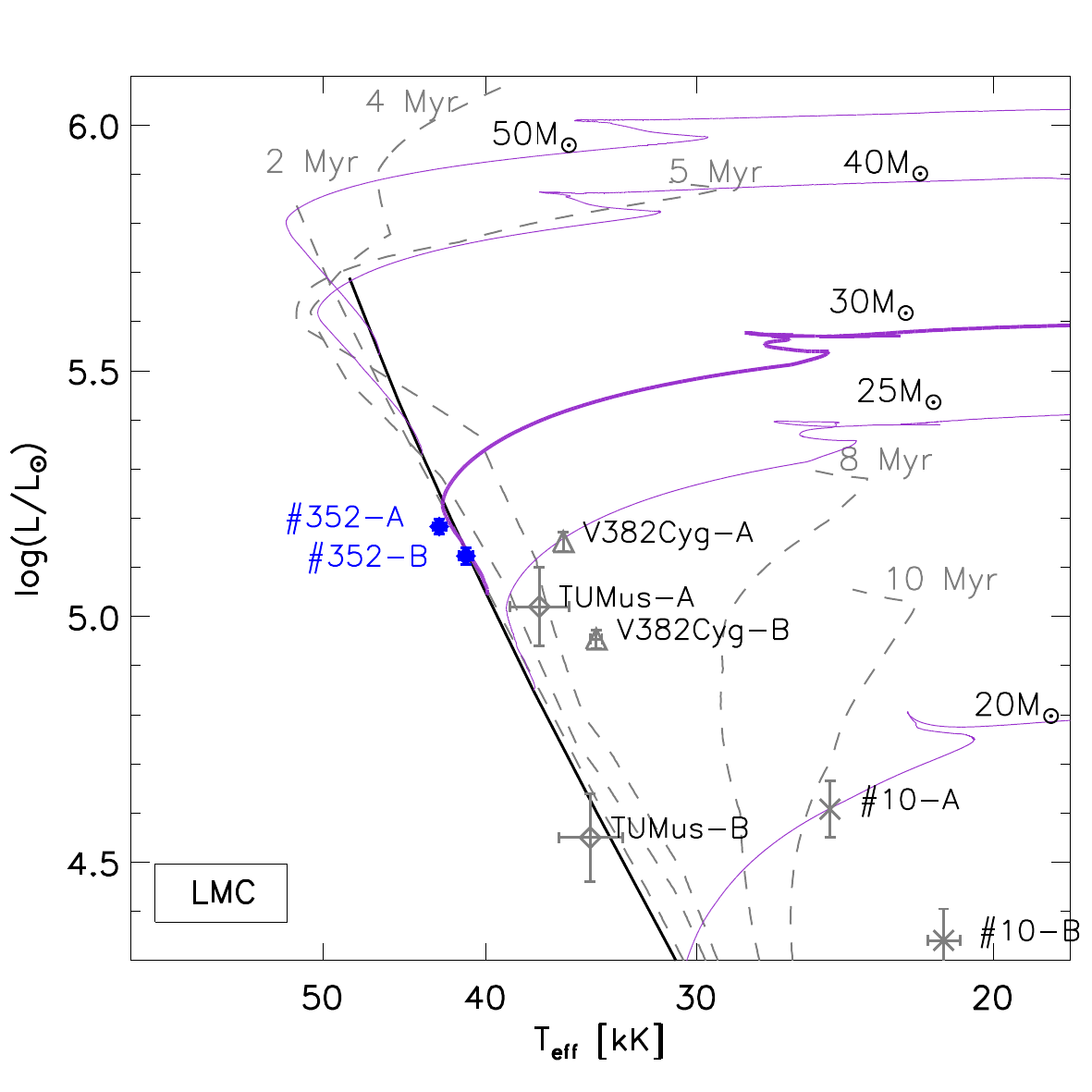}
 \caption{Location of the VFTS\,352 components on the H-R diagram. Single-star evolutionary 
 tracks (solid lines) and isochrones (dashed lines) of \citet{Brott+2011} are overlaid and 
 correspond to  initial rotation rates of 330 and 440~\kms\ for the left- and right-hand 
 side panels, respectively. The positions of the Galactic (TU~Mus, V382~Cyg) and SMC 
 (OGLE SMC-SC10 108086; \#10)  contact binaries are indicated for comparison.}
 \label{fig:evol}
\end{figure*}

\subsection{Physical properties} \label{sec:overfilling}

The simultaneous analysis of the RV and light curves of VFTS\,352 indicates that this system 
is in an overcontact configuration. While a number of massive contact binaries are known, 
i.e.\ systems where both stars just fill their Roche lobes 
\citep[e.g. MY Cam and UW CMa,][]{Lorenzo2014, Leung+1978, Antokhina+2011} only a few  
overcontact O-type systems have been identified so far:  TU Mus 
\citep[O8~V(n)z + B0~V(n); $P_{\rm orb} = 1.38$ d,][]{POG08,Sota2014} and V382~Cyg 
\citep[O7~V $+$ O7~V; $P_{\rm orb} = 1.89$~d,][]{Popper1978,DSD99} in the Milky Way, and 
OGLE~SMC-SC10~108086 \citep[O7~V + O8~V; $P_{\rm orb} = 0.88$ d,][]{Hilditch+2005} in the 
Small Magellanic Cloud. 
 
By comparing the stars' volumes with those of their Roche lobes, one can estimate the degree 
of (over)contact. For this purpose, we define the Roche volume (over)filling factor 
$f_\mathrm{L}$ as the ratio between the volume of the star and that of its Roche lobe. 
$f_\mathrm{L}$ can be approximated as 
\begin{equation}
f_\mathrm{L} = \left( \frac{R^\mathrm{mean}}{R_\mathrm{L}}\right)^3,
\end{equation}
with $R^\mathrm{mean}$, the average stellar radius and $R_\mathrm{L}$, the effective Roche 
lobe radius that we compute using \citet{Eggleton1983} approximation. According to this 
definition, both components of VFTS\,352 have $f_\mathrm{L}\approx1.29$. Stars in TU~Mus 
have $f_\mathrm{L}\approx1.30$ and 1.24 while the components of V382~Cyg and 
OGLE~SMC-SC10~108086 have $f_\mathrm{L}\approx1.1$ and 1.7, respectively.

According to their spectral classifications, the VFTS\,352 primary and secondary components 
have effective temperatures of about 42\,900~K and 40\,900~K \citep{WSSD14}. These values 
agree very well with the results of the light curves fitting --  $T_1 = 42 540\pm280$~K and 
$T_2 = 41 120\pm290$~K -- and are further consistent with the observed spectra of VFTS\,352 
(Sect.~\ref{sec:atm_models}). The good agreement between the three methods suggests robust 
effective temperature estimates.

Figure~\ref{fig:evol} compares the location of the VFTS\,352 components in the 
Hertzsprung-Russell diagram (HRD) with those of the other known overcontact binaries. It 
reveals that VFTS\,352 is  the earliest-type, hottest and most massive overcontact binary 
known to date. VFTS\,352 is also the second shortest orbital period system and the only one 
with a mass-ratio close to unity. 

\begin{figure*}
\centering
\includegraphics[width=\textwidth]{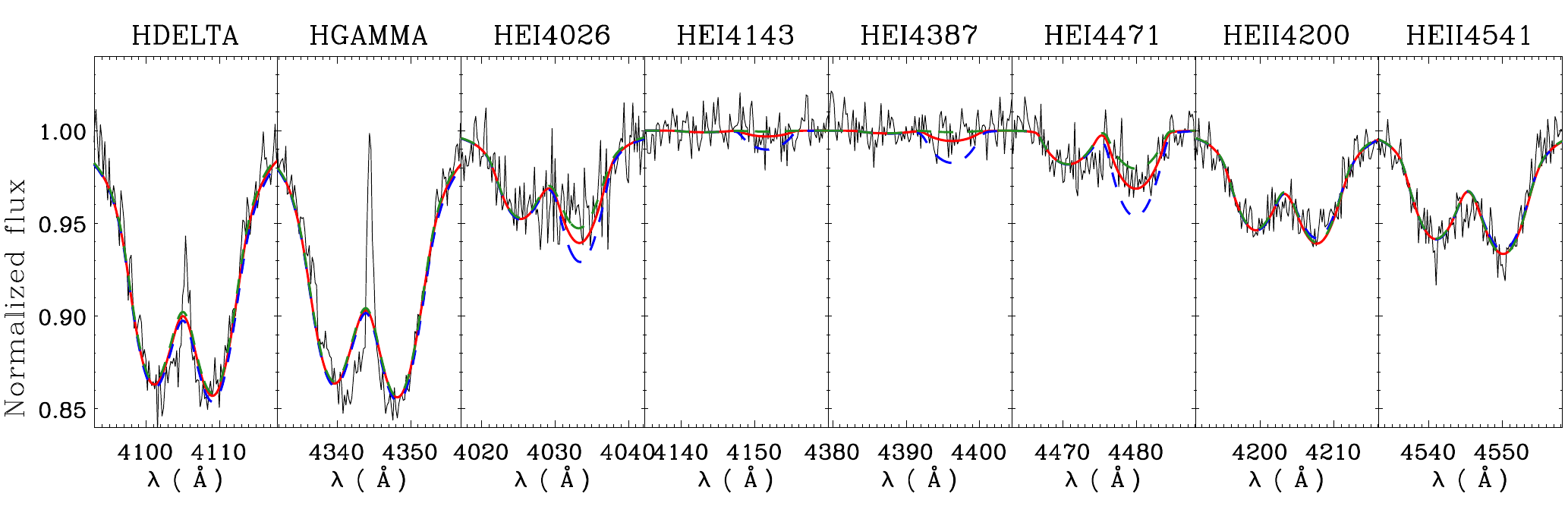}
\caption{Close up on \ion{H}{1}, \ion{He}{1} and \ion{He}{2} line profiles in the 
spectrum of VFTS\,352 ($\phi = 0.139$) compared to {\sc fastwind} models computed with 
the best fit parameters of Table~\ref{system:parameters} (plain green line) and with a 
$\pm$ 2700~K difference in the primary effective temperature $T_1$ (blue and red dashed lines).
While the green curve seems to fit the  HeI$\lambda\lambda$4026 better than the red curve, 
this helium line is a trace element in such a hot star and strongly impacted by the 
Struve Struve-Sahade effect as discussed in Section~\ref{sec:atm_models}.}
\label{fig:profile}
\end{figure*}

At 1 Myr, single stars of $28.63$ and $28.85$~\msun{} with initial spin rates 
$\approx 330$~\kms\ are expected to have effective temperatures of 39\,880 and 39\,980~K 
\citep{Brott+2011}. The VFTS\,352 components are thus hotter by about 2700 and 1100~K, 
respectively.  As a confirmation, Fig.~\ref{fig:profile} shows that the quasi-absence 
of \ion{He}{1}~$\lambda\lambda$4143, 4387 and the weak \ion{He}{1}~$\lambda$4471 in the 
primary spectrum cannot be reconciled with the temperature of 39\,880~K that is expected 
given its mass, its spin rate and the \citeauthor{Brott+2011} models. To be noted, the 
{\sc fastwind} models in Figs.~\ref{fig:spectrum} and \ref{fig:profile} are not fitted but 
serve as a  check that the light curve fitting result and, in particular, its absolute 
temperature scale, are fully consistent with the object spectrum. Interestingly, the primary 
is the most luminous and hottest component of the system, but it is also the less massive 
one. This is a clear indication that the primary has a larger helium content, thus that it 
is more evolved, than the secondary.

\subsection{Evolutionary status} \label{sec:evol}

Our photometric and spectroscopic solution shows that the system has a mass-ratio very 
close to unity. Twin binaries are expected to be common among high-mass systems with short 
orbital period \citep{Krumholz+2007}. This results from the fact that mass transfer takes 
place very early in their life. In such case, the mass transfer starts with a fast Case A, 
occurring on the thermal timescale of the components, until the system reaches a mass-ratio 
close to one. At that point, it switches to a slow Case A, which develops itself on the 
nuclear timescale \citep{KippenhahnWeigert67}. Given the relative timescales involved, it 
is more probable that we are observing VFTS\,352 during its slow Case A phase, as already 
suggesed in Sect.~\ref{sec:per}. 

There are currently no models for massive stars reliably predicting the physical properties 
of the stars {\it during} the overcontact phase.  Even though the models by \citet{deMink+2007} 
computed with the TWIN version of the Eggleton code \citep{Eggleton2006} allow for the 
formation of (shallow) contact, they do not include a realistic treatment of heat transfer 
during the contact phase. For low mass contact binaries this appears to be a crucial 
ingredient \citep{Yakut+2005}. At today's day, our results can thus not be directly 
confronted with theoretical binary computations. In an attempt to identify physical 
processes that are specific to the present configuration, we instead looked for deviations 
between the properties of VFTS\,352 and the predictions of {\it single} stellar models.

To obtain a precise interpolation of the single star evolutionary tracks of \citet{Brott+2011}, 
we use the Bayesian comparison tool BONNSAI\footnote{The BONNSAI web-service is available at 
www.astro.uni-bonn.de/stars/bonnsai.} \citep{SLdK14}. We adopt a Salpeter initial mass 
function and a flat rotational distribution as priors and we provide the temperatures, 
masses, luminosities and rotation rates that we measured. As can be seen in 
Fig.~\ref{fig:evol} (lefthand panel), no model can self-consistently reproduce all the 
observables at once: the stars appear too hot for their dynamical masses.

Comparing the Bonn's evolutionary tracks with binary results listed in the review of 
\citet{Torres+2010}, \citet{SLdK14} also found that systems with effective temperatures 
higher than 25000~K tend to be hotter than expected for their dynamical mass, although 
the difference was of the order of 1000~K only. While this could indicate a systematic 
discrepancy between observed and modelled effective temperatures, \citet{SLdK14} also 
suggested that the difference could be due to calibration issues.

In particular, we reviewed the only four O-type binaries listed in \citet{Torres+2010} 
and found that, in the original analysis of these systems, the authors used the primary 
temperature fixed from its spectral type. While this is a logical approach if the distance 
is a free parameter, it is susceptible to significant bias if a problem occurs in the adopted 
calibrations. Furthermore, the error will propagate to the secondary star properties as 
well. Indeed, the calibrations adopted in the original analyses were outdated, e.g., 
\citet{Lamers1981}, \citet{Popper1980}, \citet{Conti1973a}, \citet{Conti1973b}, etc. 
They carry a systematic shift to higher temperatures ($\sim$ 1000 K) compared to more 
recent models which take into account non-LTE, wind, and line-blanketing effects, e.g., 
\citet{Martins+2005}. For example, the V3903 Sgr system was analyzed by \citet{Vaz+1997}. 
They used \citet{Conti1973a} and \citet{Lamers1981} to fixed the primary effective 
temperature $T_1=$38000~K, for an O7V spectral type. However, an O7V has 36872 K in 
\citet{Martins+2005}, therefore, $\sim$1100 K lower. 

According to this discussion, one would need to re-analyze all hot systems listed 
in \citet{Torres+2010} using modern state-of-the-art calibration and sophisticated 
models. While this would be an important work, it lays far beyond the present paper. 
We however emphasis that, in the VFTS\,352 analysis, we do not use a fixed primary 
temperature but adjusted both $ T_1$ and $T_2$,  see Section~\ref{sec:wdc}. As a 
consequence, our analysis does not suffer from the same drawback as the historical 
analyses quoted by \citet{Torres+2010}.

Interestingly, we find that rapidly rotating single star models can reproduce the masses, 
luminosities and temperatures of the VFTS\,352 components. A good match is found adopting 
initial rotation rates of $450^{+70}_{-10}$~\kms{} (Fig.~\ref{fig:evol}, right hand panel). 
Such high rotation rates are of course ruled out by the widths of the spectral lines in 
the VLT/FLAMES data and would further not be expected given tidal locking. However, the 
good agreement between these extreme rotating models and the observed stellar parameters 
may be interpreted as evidence that the VFTS\,352 components have experienced more mixing 
than comparable single stars, possibly as a result of their binary nature.

\subsection{Signature of enhanced mixing?}

As mentioned in the previous section, the best fitting single star models are those that 
experienced enhanced internal mixing. The high temperatures and compactness of VFTS\,352 
can indeed be explained if extra helium produced in the center of the star is mixed 
throughout the envelope resulting in quasi homogeneous stars \citep{Maeder1987, Yoon+2005}. 
For the single star models of \citet{Brott+2011} that we used above, enhanced mixing is a 
consequence of rotationally induced instabilities. While the single star models are not 
appropriate for direct comparison, the good match between these fast rotating models and 
the physical properties of the VFTS\,352 components may be a signature that the stars have 
experienced additional mixing compared to similar single star with a 330~\kms\ rotation 
rate, even if the nature of the additional mixing processes in VFTS\,352 may differ from 
those implemented in single star models.   

Stars in close binary systems are generally thought to experience enhanced mixing in 
comparison with a single stellar models with the same birth rotation rate 
\citep[e.g.,][]{de-Mink+2009,Song+2013}. For example, (1) tidal locking of the rotation 
period of the outer stellar layers with the orbital period creates internal shear. The 
shear can persist as the star evolves and its internal structure slowly changes. As the 
tides force the stars to stay in synchronous, (2) angular momentum from the orbit is 
converted into the spin angular momentum of the stars. For single stellar models higher 
angular momentum implies more internal mixing. If this holds true for binary systems, 
it would imply enhanced mixing.  Finally, (3) the deformation of stars that are very 
close to filling their roche lobe may also play a role. In single stars the deviation 
of spherical symmetry is the seed for the Von Zeipel effect and the Eddington-Sweet 
circulations. For close binary stars the symmetry around the rotation axis is also 
broken, adding further possibilities for internal processes to enhance mixing.

It is unclear how these processes (and possible additional processes not listerd here) 
operate and possibly interplay with each other. Attempts to account for (some of) this 
additional physics were made by \citet{de-Mink+2009} and later independently by 
\citet{Song+2013}. Both groups indeed predict that stars in compact binaries experience 
more mixing than single stars with identical birth rotation rate. This is consistent 
with our finding that the best matching single star models have higher initial rotation 
rates than expected for VFTS\,352. 

The models by \citet{de-Mink+2009} further  predict enhancements of the surface with 
CNO burning products. The rapidly rotating single star models that reproduce our 
observations imply a surface helium mass fraction of 39 and 32\%\ for the primary 
and secondary components. These indirectly inferred abundances are consistent with 
the detection limits obtained from our spectra.

\subsection{Alternative scenarios without enhanced mixing}

An alternative explanation for the high temperatures and increased helium surface abundance 
that does not invoke enhanced mixing involves stripping of a star from its enveloppe. This 
exposes the deeper layers which show signatures of the nuclear burning in the center of the 
convective core. In absence of extra mixing one needs to strip the star(s) down to the 
layers that were once part of the convective core.  The removal of a star's outer layers 
can be achieved either through stellar winds or by Roche lobe overflow and subsequent mass 
transfer. Here we briefly discuss these scenarios.

While stellar winds can play a major role in the advanced evolutionary stages of massive 
stars, we expect them to have had a negligible effect for the relatively unevolved 
components of VFTS\,352. There are no significant wind spectral signatures in the 
spectrum of VFTS\,352 \citep{WSSD14}. The reduced metallicity of the LMC also implies 
that the role of stellar winds in this early stage is neglible. We thus conclude that 
stripping by stellar winds alone is unable to remove a substantial part of the envelope.

The second scenario concerns Roche lobe overflow by the initially most massive star and 
(partial) accretion by the companion. Allthough some mass loss due to Roche lobe overflow 
seems likely, it is hard -- within our current understanding of binary interaction -- to 
reconcile the VFTS\,352 observables using mass transfer alone (i.e.\ without invoking 
extra mixing).  To explain the surface helium enrichment of the primary star --  39\%\ 
by mass -- in absence of extra mixing, the star needs to shed about half of its total 
mass. This implies an extreme initial mass ratio  $M_2/M_1 \lesssim 0.5$, with the exact 
value depending on how much the mass gainer accreted. Mass transfer in systems with 
such extreme mass ratios is expected to be highly non-conservative due to the thermal 
expansion of the mass gainer \citep{Pol94,Wellstein+2001,deMink+2007}. Also the current 
short orbital period of VFTS\,352 may be seen as an indication of a prior non-conservative 
evolution that has resulted in a shrinkage of an originally wider orbit.  

Yet, this picture is very hard to reconcile with the inferred helium surface enrichment 
of the secondary -- 32\%\ by mass fraction -- which suggests that the secondary has 
accreted at least some of the primary enriched material. A simple back-of the enveloppe 
calculation indeed shows that, assuming a primary enriched material with a 39\% He-mass 
fraction at most, the secondary star should have accreted at least 6~\msun\ in order to 
increase the He-mass fraction of its enveloppe from 26\%\ to 32\%. This contradicts the 
non-conservative nature of the scenario implied by the very low initial mass-ratio and 
current short orbital period.  Reversing the role of the primary and secondary in this 
scenario does not help, hence mass transfer alone is unable to provide a satisfactory 
explanation either.

In view of the current state of the models, it seems that neither stellar winds nor Roche 
lobe overflow, acting alone, are  sufficient mechanisms to provide a simple and natural 
explanation for the inferred properties of VFTS\,352. We thus suggest enhanced internal 
mixing, either as the sole culprit or in combination with the processes above, as the 
most probable explanation for their current properties.

\subsection{Age constraints}
If the VFTS\,352 components are evolving quasi-homogeneously under the combined effects 
of rotation and tidally-induced enhanced mixing, the position of the stars in the HRD 
would then remain close to the  zero-age main sequence for the first few Myr of their 
life \citep{Brott+2011}. Given our constraints on the temperature and luminosities, 
we obtain age estimates of 3.4$\pm$0.2 and 2.4$\pm$0.2~Myr for the primary and the 
secondary. Both the higher helium mass fraction of the primary and the different age 
estimates provide indications of a previous mass-transfer event. 

The age that we infer opens up to the possibility that the system was not born at its 
current location but may have migrated from one of the young 30 Doradus clusters after 
receiving an impulse through dynamical interaction \citep[e.g.][]{Fujii2011Sci}. The 
systemic radial velocity, $\gamma=262.8\pm1.2$~\kms, is in overall agreement with that 
of the 30 Doradus region and does not indicate that the system is a runaway star along 
the line of sight. We inspected the OGLE data and, while the dispersion of the measurements 
is large, a significant proper motion of the order of $-0.8\pm0.3$ milli-arcsec yr$^{-1}$ 
may be present along the  declination axis. While error bars are large, such a proper 
motion allows for the possibility that the system has been ejected from the NGC~2070 
association or from R136, the very massive cluster at its core. An ongoing HST proper 
motion program (GO 13359, PI: Lennon) will soon be able to confirm the tentative 
proper-motion detected in the OGLE data.

\subsection{Final fate} \label{sec:fate}

Two possible outcomes of the overcontact configuration of VFTS\,352 can be envisioned, 
depending on whether the stars will evolve  redward or blueward in the HRD. In the 
first scenario, the build-up of an internal chemical gradient will reduce the efficiency 
of the rotationally- and (possibly) tidally-induced mixing over time. Stars will then 
switch to a more standard evolution scenario in which secular expansion will push the 
VFTS\,352 components into deeper contact and the system will eventually merge.

While the merging process itself is poorly understood, it may produce an intermediate 
luminosity transient \citep{Soker2006}, such as the one associated with the merger of 
the lower-mass system V1309 Sco  \citep{Tylenda+2011}. The likely outcome of the merging 
process is a rapidly rotating and possibly magnetic single star \citep{Ferrario+2009, 
Langer2012, de-Mink+2013, de-Mink+2014}. According to the models by \citet{Yoon+2006} 
and \citet{Brott+2011}, such a fast rotating massive star may itself experience 
significant mixing induced by rotation. A similar system at low metallicity might even 
retain enough angular momentum to fulfill the requirements of the collapsar scenario 
\citep{Woosley1993} for the progenitors of long GRBs as suggested by \citet{Yoon+2005} 
and \citet{Woosley+2006}.

In the second scenario, the internal mixing is such that the stars would pursue 
(quasi) chemically homogeneous evolution. They would become hotter and more luminous, 
but would remain compact. The system may thus avoid the coalescence phase described 
above and the stars would retain their tidally-locked rapid rotation rate throughout 
their main sequence lifetime \citep{de-Mink+2009}. Once most of the stellar hydrogen 
will have been burned, the VFTS\,352 components are then expected to enter a Wolf-Rayet 
phase and to develop strong stellar winds. 

Tides can in principle counteract the spin down induced by  mass-loss  and preserve the 
rapid rotation. This however depends on uncertain aspects such as how much the orbit 
widens during this phase and how efficient the tides are. If the components in VFTS 
352 can retain their high spins until the end of their lives they would be interesting 
candidates as progenitors of long Gamma-ray bursts, despite the relatively high-metallicity 
environment provided by the LMC.    

Regardless of the evolution of the stellar spin,  a scenario in which VFTS\,352 avoids 
coalescence implies two massive stars whose cores are enlarged as a result of mixing. 
In such a configuration, the most likely final outcome of is the formation of two 
stellar mass black holes. This makes VFTS\,352 also interesting as a potential progenitor 
of binary black hole merger and gravitational wave source.

\section*{ACKNOWLEDGMENTS}
This work is based on data obtained at the European Southern Observatory under 
program IDs.\ 182.D-0222, 090.D-0323, and 092.D-0136. The authors are grateful to 
the referee for comments that help improving the manuscript, to N.R. Walborn for 
fruitful discussions and to  J. Pritchard and the ESO staff for their support during 
the observing campaigns. LAA acknowledges support from the Funda\c{c}\~{a}o de 
Amparo \`a Pesquisa do Estado de S\~{a}o Paulo - FAPESP (2013/18245-0 and 2012/09716-6); 
SdM, from a Marie Sklodowska-Curie Reintegration Fellowship (H2020-MSCA-IF-2014, project id 661502) 
awarded by the European Commission; J.M.A., from the Spanish 
Government through grant AYA2013-40\,611-P; R.H.B., from FONDECYT No 1140076. This work 
has been partly supported by the Polish National Science Centre grant No.\ DEC-2011/03/B/ST9/02573.


\end{document}